# A Novel MDP Based Decision Support Framework to Restore Earthquake Damaged Distribution Systems

Ebru AYDIN GOL,
Dept. of Computer Eng.,
METU,
Ankara, Turkey

Burcu GÜLDÜR ERKAL,
Dept. of Civil Eng.,
Hacettepe University,
Ankara, Turkey

Murat GÖL,
Dept. of Electrical & Electronics Eng.,
METU,
Ankara, Turkey

*Abstract* — Electric power network expanded rapidly in recent decades due of the excessive need of electricity in every aspect of life, including critical infrastructures such as medical services, and transportation and communication systems. Natural disasters are one of the major reasons of electricity outage. It is extremely important to restore electrical energy in the shortest time possible after a disaster. This paper proposes a decision support method for electric system operators to restore electricity to the critical loads in a distribution system after an earthquake. The proposed method employs Markov Decision Process to find the optimal restoration scheme based on the Probability of Failure of critical structures determined by using the Peak Ground Acceleration values recorded by observatories and earthquake research centers during earthquakes.

## I. Introduction

Fast restoration of electricity after a disaster is extremely important, since modern life, as we know it, relies on the presence of electricity. Black-start is a hard problem to solve even if the considered system is not structurally damaged. However, after a disaster, it is highly probable that the field instruments may be destructed. Especially overhead lines and transformers at urban areas are prone to get damaged during an earthquake. In this study, a decision support method for distribution systems hit by an earthquake is developed. The proposed method determines the best strategy to energize the power lines based on the probability of exceeding the damage state of failure (Probability of Failure - $P_F$) obtained by using the recorded peak ground acceleration (PGA) values. The method determines the optimal strategy using Markov decision process (MDP).

Power system restoration and disaster management have been studied widely in various aspects in the last decade [1] – [9]. [1] and [2] aim to develop recovery plans for expected disasters and outages. [3] evaluates use of micro-grids for fast restoration of power systems. [4] – [7] propose online disaster management methods with the help of field information provided by sensors. However, those methods do not consider destruction of field components, such that they assume none of the transformers and transmission lines are damaged. Moreover, they require deployment of field sensors. Whereas, the proposed decision support mechanism for system restoration only requires basic SCADA data and $P_F$, which is determined by using the real time earthquake data. In [8] and [9] planning strategies to improve the resilience of the power systems are presented. In this paper, rather than a planning strategy, a decision support method that will be utilized in the presence of a disaster is proposed. This method considers the properties of the realized earthquake and the existing infrastructure of the distribution system.

In this paper, a novel framework based on MDPs to generate optimal restoration strategy, i.e. the strategy with minimum expected restoration time, for a medium voltage (MV) distribution system after an earthquake is presented. The proposed method models the restoration of electric distribution system as an MDP, such that a state of the MDP represents the overall information about the network. In particular, an MDP state shows the physical status of each branch, such that a branch is in either good (energized), damaged (broken/cannot be energized) or unknown condition. An action represents the set of branches that can be energized in a state and the transition probabilities are computed according to the recorded PGA data. The PGA data is recorded by observatories and earthquake research institutes, and the data is published after an earthquake is experienced. The $P_F$ values can be calculated easily once the PGA data is received.

The proposed model reduces the synthesis of optimal restoration strategy to optimal policy synthesis problem for MDPs. Moreover, the new modeling formalism supports incorporation of further constraints into the synthesis of the restoration strategy. The method uses real time data based on $P_F$ to achieve its minimal restoration time goal by avoiding paths with the highest failure probability. The proposed method does not require any additional infrastructure. The MDP formulation enables assessment of each sequence of actions; therefore, it provides an optimal restoration strategy.

## II. Proposed Method

### A. Fragility Analysis

Earthquakes, due to their randomness of the return period and the shaking intensity, are one of the most unpredictable natural hazard types. Therefore, it is required to perform seismic performance assessment of the existing structures in

This work is supported by Scientific and Technological Research Council of Turkey (TUBITAK) under project number 118E183, and the European Union's Horizon 2020 research and innovation program under the Marie Sklodowska-Curie grant agreement No 798482.

order to compute the safety against seismic hazard. Fragility curves are used to represent the seismic risk exposure based on the risk and safety assessment of structural components. These curves show the probability of exceeding a structural performance level with respect to increasing earthquake intensity level [10]. There have been an increasing number of studies in the past two decades that focus on the use of fragility curves for estimating seismic damage levels after natural hazards [11] – [14]. In this paper, the structural performance levels are represented by the damage states and the earthquake intensity levels are shown in terms of PGA values. The proposed method supplies $P_F$ values, which are calculated based on PGA, to an MDP, in order to represent the possible current status of the field structures, i.e. overhead lines and transformer sub-stations.

Fragility curves for varying damage levels represent the seismic damage vulnerability of structures. To obtain these curves, first, structures are modeled with software, and are analyzed under simulated earthquake loads. The fragility curves are then obtained by importing the obtained seismic response output into the fragility function.

In order to formulate the analytical fragility function, it is required to investigate the probability of seismic demand (DM) exceeding a limit-state (LS) for varying seismic intensity measure (IM) values by using the "Total Probability Theorem" [15]. The results of this investigation demonstrate two major elements in seismic risk analysis; local seismicity, and the probability of seismic damage and structural fragility. Local seismicity is generally represented by readily available seismic hazard curves whereas the probability of seismic damage and structural fragility are denoted by fragility curves. To plot the mentioned fragility curves, a closed-formed solution for the fragility function is obtained by forming a relationship between motion intensity and the seismic demand [16]. Fig. 1 displays a sample fragility curve collection that represents mean annual frequency (MAF) of seismic hazard estimated for a range of ground motion intensities with different magnitudes and epicentral distances versus IM. Each curve corresponds to a different structural performance.

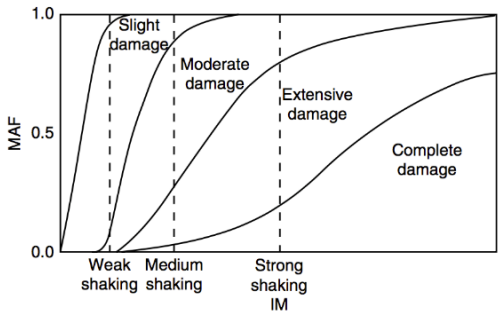

Fig. 1. Sample fragility curves for slight to complete damage states [10].

Note that only the fragility curves related to the complete damage (collapse) level that corresponds to the loss of functionality is considered with the assumption of definite power outage at this damage level. The effects of the preceding damage levels are considered to be benign on the electric distribution system. Thus, only the annual probability of exceeding the damage level of failure ($P_F$) is computed and plotted. At the same time, since the investigated structures are assumed to be not located in liquefiable or landslide zones, the IM is selected as PGA, which is equal to the amplitude of the largest absolute acceleration recorded on an accelerogram at a site during a particular earthquake. The seismic fragilities of the investigated structures are computed and plotted accordingly. Data published by observatories and earthquake research institutes immediately after an earthquake can be used to determine the PGA values.

*B. Markov Decision Process (MDP)*

MDP is a mathematical framework used to make decisions in probabilistic environments [17]. An MDP is composed of states, actions and transition probabilities that capture the dynamics of the actual system. An MDP policy determines the action to take in a state. In its simplest form, the goal of the Markov decision problem is to synthesize the policy optimizing the given criteria. This problem has been studied from various aspects such as maximizing the probability to reach a set of states, minimizing the expected time to reach a state [17], or minimizing the cost under additional constraints such as safety or formal specifications [18].

An MDP is a tuple $M = (S, A, p, c)$, where $S$ is a set of states, $A$ is a set of actions, $p$ is a state transition probability function, i.e., $p(s'|s, a)$ is the probability of the system to transit to state $s'$ when action $a$ is applied in state $s$, and $c: S \to \mathbb{R}$ is a state cost function that assigns a cost to each state. In a state $s$, actions from a certain set can be applied, which is denoted by $A(s)$. A deterministic MDP policy $\pi: S \to A$ with $\pi(s) \in A(s)$ determines the action to be applied in a state-$s$. A value function $v_\pi: S \to \mathbb{R}$ represents expected cost obtained following policy $\pi$ from each state in $S$. The total expected cost incurred until a target set $G \subseteq S$ is reached is computed with the following value function:

$$v_\pi(s) = 0 \quad \text{if} \quad s \in G$$
$$v_\pi(s) = c(s) + \sum_{s' \in S} p(s'|s, \pi(s)) \cdot v_\pi(s') \quad (1)$$

Objective of the MDP is to synthesize an optimal policy. For a given state $s_0$, the optimal policy $\pi^*$ is the policy that minimizes the value function: $\pi^* = \underset{\pi}{\text{argmin}}\, v_\pi(s_0)$.

*C. The Proposed Decision Support Strategy*

The restoration of earthquake damaged electric distribution system is modeled as an MDP. Note that the paper focuses on the MDP based decision support framework; therefore, the problem is simplified using the following assumptions.
1. Electrical load forecasts and generation forecasts of the distributed energy sources (DERs) are available.
2. A power flow analysis tool runs to check if the bus voltages are within permissible limits.
3. Low voltage system cannot be controlled.
4. MV circuit breaker status can be monitored and controlled remotely via the SCADA system.

In the proposed model, a state of an MDP represents the information about all branches of the network. A branch of the network can be in one of the following states: energized ($E$), damaged ($D$, cannot be energized) or unknown ($U$). The unknown state implies that the branch is not attempted to be energized. Hence, the corresponding circuit breakers are open, and the physical status of the branch (damaged or healthy) is unknown to the decision maker (system operator). To exploit the available information after the disaster, the $P_F$ data is used to compute the probability of being damaged.

A restoration action can be applied to a branch if it is at the unknown state ($U$) and connected to an energized branch ($E$) or a source. The state of a branch can change to either $D$ or $E$ from $U$ if the corresponding action is applied, i.e., the corresponding circuit breakers are closed. However, the state of a branch cannot change if it is $D$ or $E$. Thus, there is no admissible action for the branch in $D$ or $E$.

The MDP model $M = (S, A, p, r)$ of the restoration process is defined as follows. Let $m$ be the number of branches. The combination of all possible states of all branches is the state space of the model:

$$S = \{s_0, s_1, \dots, s_N\}$$
$$\text{where } s_i = [s_i^0, s_i^1, \dots, s_i^{m-1}] \text{ and } s_i^k \in \{U, D, E\} \quad (2)$$

As seen in (2), there can be at most $3^m$ states, i.e. $N \leq 3^m$. However, a large fraction of those states represents infeasible system states, e.g., a branch cannot be $E$ if it is not connected to a source or an energized branch. Such states are never constructed. Hence the cardinality of $S$ is much less than its theoretical bound $3^m$ in practice, thanks to the employed iterative construction method with reachability analysis.

In a network, restoration action can be applied to multiple branches simultaneously if those are electrically distant enough such that the transients emerging after the closure of the circuit breakers do not cause significant effects. In this work, if none of the ends of branches are connected, then they are considered as electrically distant for the sake of simplicity.

Action $a \in A$ of model $M$ is the set of branches to be energized simultaneously and is defined as a subset of the branch indices $A^b = \{0, 1, \dots, m-1\}$, i.e., $a \subseteq A^b$ and $A$ is the powerset of $A^b$. The restoration action to branch $j$ will be applied when a model action $a \in A$ with $j \in a$ is applied in the model level. During the model construction, the set of feasible actions (i.e., $A(s)$) is defined with respect to the system structure and constraints. If two branches have a common end, they are said to be connected. The branches connected to branch $j$ is denoted by $C(j)$. Let $A^b(s_i) \subseteq A^b$ denote the feasible branch actions for a state $s_i = [s_i^0, s_i^1, \dots, s_i^{m-1}] \in S$. The restoration action for branch $j$ is feasible, i.e., $j \in A^b(s_i)$ if the following conditions hold; (i) $s_i^j$ is $U$ (the status of branch $j$ is $U$), (ii) $s_i^k$ is $E$ for some $k \in C(j)$ (branch $j$ is connected to an $E$ branch) or $s_i^j$ is connected to a source.

If $A^b(s_i) = \emptyset$, then there is no feasible action. In other words, the system operator cannot change the state of the system by closing circuit breakers. This occurs when no branch in U state is connected to a source or an energized branch. These states are called terminal and the set of terminal states is denoted by $S^T = \{s \mid A^b(s) = \emptyset\}$.

In order to restore energy in a distribution system, it is sufficient to create a spanning forest of energized branches with at least one power source (transmission grid or DER) connected to each tree. Note that, for operational purposes, meshed structures are avoided in distribution systems. Therefore, the resulting strategy should not include a loop of energized branches. The mesh constraint and the electrical distance constraint are integrated to the model $M$ via the feasible actions $A(s_i)$ such that a subset $a$ of $A^b(s_i)$ is added to $A(s_i)$ only if the following conditions are satisfied: (i) energizing branches in $a$ does not generate a loop of energized branches, (ii) all branches in $a$ are electrically distant.

The proposed method guarantees that all system properties and constraints are integrated into model $M$ via the feasible action set definition ($A(s)$). Consequently, when the distribution system is in a state $s$, any restoration action from $A(s)$ can be applied without further computation. Furthermore, applying an action $a \notin A(s)$ is either infeasible or it can cause violation of the above-mentioned constraints.

Next the computation of the transition probabilities from $P_F$ data as is explained. In state $s = [s^0, s^1, \dots s^{m-1}]$, when action $a \in A(s)$ is applied, the probability of transitioning to a state $t = [t^0, t^1, \dots t^{m-1}]$ satisfying (3) is given in (4). Essentially, by $A(s)$ definition $s^i = U$ for all $i \in a$ and the status of these branches change to $D$ or $E$ according to $P_F$. The transition probability that does not satisfy (3) is 0 as those states are not one step reachable from $s$ under the control action $a$.

$$t^i = s^i \text{ if } i \notin a, \qquad t^i \neq s^i \text{ if } i \in a \quad (3)$$

$$p(t \mid s, a) = \prod_{i \in a} \begin{cases} P_F(i) \text{ if } t^i = D \\ 1 - P_F(i) \text{ if } t^i = E \end{cases} \quad (4)$$

In this study, the goal is to minimize the restoration time. This optimization criterion is integrated to the MDP problem via the minimization of the time to reach to a terminal state $s \in S^T$, which is defined via the following cost function:

$$c(s) = \begin{cases} 0 \text{ if } s \in S^T \\ 1 \text{ otherwise} \end{cases} \quad (5)$$

The value function computed for (5) gives the expected number of steps to restore energy. Hence, the value of the optimal strategy is the optimal expected restoration time.

*Remark1:* Given the cost function (5), the optimal strategy $\pi^*$ ensures that all feasible actions will be applied in the minimum number of steps. This criterion guarantees that a proper superset of $\pi^*(s)$ does not belong to $A(s)$ for any $s \in S$. This property is exploited in the model construction to reduce the model size. In particular, first, the set of all feasible control actions $A(s)$ is computed as explained previously. Then, the actions $a' \in A(s)$ whose superset $a$, $a' \subset a \in A(s)$, is also included in $A(s)$ is removed:

$$A'(s) = \{a \in A(s) \mid a \cap a' \neq a \text{ for each } a' \in A(s) \text{ such that } a \neq a'\} \quad (6)$$

The reachability analysis is performed according to reachable states computed from $A'(s)$. Note that the optimal strategy and the optimal cost does not change when $A'(s)$ is used instead of $A(s)$ since for the optimal strategy $\pi^*$ computed without the simplification it is guaranteed that $\pi^*(s) \in A'(s) \subseteq A(s)$ for each state $s$.

*Remark2:* In addition to the constraints 1 and 2 listed above for the construction of feasible actions, the limited capacity of a DER should be considered as a third constraint. In particular, energizing branches in $a$ should not create an energized tree that is connected to a DER with the total demand exceeding the capacity of the DER. The constraint can easily be integrated in the model construction phase. However, in the policy synthesis phase, the infeasibility of energizing a $U$-branch due to the DER capacity should be considered. This situation should be avoided as long as the branch can be energized via an alternative policy. Therefore, it can be achieved via a reach-avoid specification. Due to the space limit and for the sake of presentation, it is assumed that each DER has sufficient capacity to supply all loads. The method presented in this paper generates a suboptimal strategy when this assumption is violated.

*Example.* The model construction method is illustrated on a toy system with 6 buses shown in Fig. 2. Node 1 is connected to the transmission substation and node 6 is connected to a DER. Before the earthquake the circuit breakers between buses 4 and 5 are open. After the earthquake, system operator may utilize this branch in order to gain a leverage to supply energy to the customers. Therefore, the proposed method considers all available branches of the system, independent of their initial status. After the earthquake, initially, all the breakers are open and the states of the branches are unknown and $P_F(i) = 0.2$ for each branch i. Thus, $s_0 = [U, U, U, U, U, U]$ is the initial state, where the branches are defined in Fig. 2.

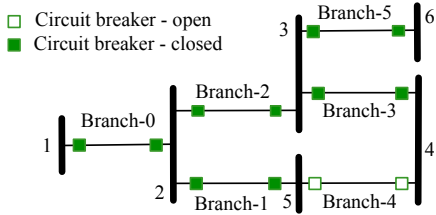

Fig. 2. Sample system with circuit breaker statuses.

As only branch-0 and branch-5 are connected to the energy sources and no branch is energized, $A^b(s_0) = \{0,5\}$. Since branches 0 and 5 are electrically distant, energizing these branches does not create a mesh or overload the DER, the restoration action can be applied simultaneously and $A(s_0) = \{\{0\}, \{5\}, \{0,5\}\}$. As $\{0\}$ and $\{5\}$ are proper subsets of $\{0,5\}$, the simplified action set defined in *Remark1* is $A'(s_0) = \{\{0,5\}\}$. The transition probabilities for the reachable states are computed according to (4) for $\{0,5\}$ and new states are added to $S$. The states that are reachable from $s_0$ under the control action $\{0,5\}$ and their transition probabilities are shown in (7).

$$p(s_1 = [E, U, U, U, U, E]|s_0, \{0,5\}) = 0.64$$
$$p(s_2 = [E, U, U, U, U, D]|s_0, \{0,5\}) = 0.16$$
$$p(s_3 = [D, U, U, U, U, E]|s_0, \{0,5\}) = 0.16 \quad (7)$$
$$p(s_4 = [D, U, U, U, U, D]|s_0, \{0,5\}) = 0.04$$

After this computation, 4 new states are added to $S$. Note that $s_4$ is a terminal state since $A^b(s_4) = \emptyset$. The computation continues iteratively, i.e., action sets and transitions are computed for each new state. If $A(s_0)$ was used instead of $A'(s_0)$, then 4 more new states would be added. However, according to the given criteria $\pi^*(s_0) = \{0,5\}$ (rather than $\{0\}$ or $\{5\}$) independent of the rest of the model, and those states that are not added are not reachable when $\pi^*$ is applied. This simplification reduces the model size, hence the computation time, significantly. For this toy example, the numbers of states are 58 and 188, with and without the simplification, respectively. Although the model size is reduced, the expected restoration time (2.6588), which is the value of the optimal strategy, are the same, as expected. ∎

The proposed approach can be adapted to minimize the restoration time to some particular customers (e.g. a hospital or a data center) instead of the whole network. Let $i$ be the index of the bus supplying this customer, and let $S^{T,i}$ be the set of states that are reachable from the initial state and, $S^{T,i} = \{[s^0, s^1, ..., s^{m-1}] \in S \mid s^j$ is $E$ for a branch $-j$ adjacent to bus $-i\}$ (bus-$i$ is energized). When $S^{T,i}$ is used in (5) instead of $S^T$, the resulting policy minimizes the expected restoration time for branch-$i$. Once the branch is energized (or found to be damaged), the method can be applied again to synthesize a strategy for the remaining parts.

### III. SIMULATIONS AND RESULTS

In this section, a simple system, shown in Fig. 3, is considered for 4 different $P_F$ scenarios, to validate the proposed method. Note that, in all simulations, the considered systems are hit by an earthquake, and hence experiencing an energy shortage.

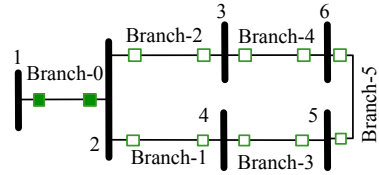

Fig. 3. 6-bus test system.

Bus 1 is the MV side of the power transformer. When all the circuit breakers are open, intuitively, the restoration strategy first closes the circuit breakers of branch-0, and then needs to choose closing the circuit breakers of either branch-1 or branch-2. As soon as the chosen branch is energized, then the restoration actions can be applied in both paths simultaneously (e.g. branches 2 and 3 can be energized after branch-1 is energized) thanks to the previously stated assumption of having enough electrical distance if two branches are not adjacent. To restore the energy in the minimum amount of time, it is clear that the breakers along the branches with less probability of destruction should be closed first for this simple example. As explained in the paper, the aforementioned constraints and the optimality criteria are integrated to the MDP problem and the optimal MDP strategy minimizes the expected restoration time considering $P_F$ for the whole network. To illustrate, four earthquake scenarios are considered and the corresponding $P_F$ data is show in Table I. Note that branch-0 is skipped, as it is assumed to be not damaged to continue operation.

*Scenario 1:* The optimal strategy first produces action sequence $\{2\}\{1,4\}\{3\}$. This sequence will be applied unless a damaged branch is observed. If no damaged branch is observed, the corresponding MDP state sequence is:

$$[UUUUU] \to_{0.6}^{\{2\}} [UEUUU] \to_{0.18}^{\{1,4\}} [EEUEU] \to_{0.6}^{\{3\}} [EEEEU]$$

where $\to_p^I$ denotes that the circuit breakers for branches in set I will be closed and the probability to see the state on the right-hand side after this action is $p$. If a damaged branch is observed, the action sequence will be updated according to the optimal MDP strategy. Note that, branch-0 is avoided in MDP state sequence, since it is assumed to be not damaged.

TABLE I. $P_F$ VALUES FOR THE SCENARIOS

| Scenario | $P_F$ | | | | |
|---|---|---|---|---|---|
| | Branch-1 | Branch-2 | Branch-3 | Branch-4 | Branch-5 |
| 1 | 0.7 | 0.4 | 0.4 | 0.4 | 0.4 |
| 2 | 0.4 | 0.7 | 0.4 | 0.4 | 0.4 |
| 3 | 0.4 | 0.4 | 0.4 | 0.7 | 0.4 |
| 4 | 0.4 | 0.4 | 0.7 | 0.4 | 0.4 |

Assume that only branch-1 is damaged, then the action sequence will be updated after {1,4} is applied. The strategy will produce sequence {5}{3} and the following trace will be observed.

$$[UUUUU] \to_{0.6}^{\{2\}} [UEUUU] \to_{0.42}^{\{1,4\}} [DEUEU] \to_{0.6}^{\{5\}} [DEUEE] \to_{0.6}^{\{3\}} [DEEEE]$$

*Scenario 2:* The optimal strategy produces the action sequence {1}{2,3}{4}. Note that the only difference between first and second scenarios is the failure probabilities for branch-1 and branch-2. The optimal strategy first energizes the one with higher probability of success.

*Scenario 3:* The produced action sequence is {1}{2,3}{5}.

*Scenario 4:* The produced action sequence is {2}{1,4}{5}.

In scenarios 1 and 2, the optimal action can be decided according to the branches connected to the source. Whereas, in scenarios 3 and 4, the optimal action changes according to the $P_F$ data of other branches that are not directly connected to the source. The example, while being extremely simple, illustrates that the optimal strategy obtained from the MDP problem considers future steps as well. For example, in scenarios 3 and 4, the optimal strategy avoids branch-4 and branch-3 respectively, as there is an alternative way to energize the whole network with less expected time (via branch-5).

## IV. CONCLUSIONS

This paper proposes an MDP based decision support method for systems that are experiencing interruption after an earthquake. The proposed method employs $P_F$ values in order to determine probability of failure of a distribution system component or a structure that is located close to the distribution system. $P_F$ is determined using real time data recorded during the earthquake by observatories and research centers. Note that, the proposed method does not require any additional infrastructure.

The proposed method aims to find the best energy restoration strategy by minimizing the probability of unsuccessful actions. As feedback on the topology of the system is gathered from the field, the proposed method updates the solution.

The formed MDP problem is solved with a reasonable duration. As the operator receives feedback from the field, the updated optimum strategy can be found by updating the MDP state without considering the whole problem. Note that, the system operator runs the proposed method after the interruption to restore energy to customers, and hence no real time solution, which minimizes the total restoration duration, is required but rather a fast-enough solution is sufficient.

## V. REFERENCES

[1] A. Arab, A. Khodaei, S. K. Khator, K. Ding, V. A. Emesih and Z. Han, "Stochastic Pre-hurricane Restoration Planning for Electric Power Systems Infrastructure," in *IEEE Transactions on Smart Grid*, vol. 6, no. 2, pp. 1046-1054, March 2015.

[2] F. Qiu and P. Li, "An Integrated Approach for Power System Restoration Planning," in *Proceedings of the IEEE*, vol. 105, no. 7, pp. 1234-1252, July 2017.

[3] Z. Zhao and B. T. Ooi, "Feasibility of fast restoration of power systems by micro-grids," in *IET Generation, Transmission & Distribution*, vol. 12, no. 1, pp. 126-132, 1 2 2018.

[4] A. Golshani, W. Sun, Q. Zhou, Q. P. Zheng and J. Tong, "Two-Stage Adaptive Restoration Decision Support System for a Self-Healing Power Grid," in *IEEE Transactions on Industrial Informatics*, vol. 13, no. 6, pp. 2802-2812, Dec. 2017.

[5] N. Ganganath, J. V. Wang, X. Xu, C. T. Cheng and C. K. Tse, "Agglomerative Clustering Based Network Partitioning for Parallel Power System Restoration," *IEEE Transactions on Industrial Informatics*, vol.14, no. 8, pp. 3325-3333, Aug. 2018.

[6] L. H. T. Ferreira Neto, B. R. P. Júnior and G. R. M. da Costa, "Smart Service Restoration of Electric Power Systems," *2016 IEEE Power and Energy Society General Meeting (PESGM)*, Boston, MA, 2016, pp.1-5.

[7] M. Ostermann, P. Hinkel, D. Raoofsheibani, W. H. Wellssow and C. Schneider, "A minimum-regret-based optimization approach for power system restoration in EHV grids," *2017 IEEE Power & Energy Society General Meeting*, Chicago, IL, USA, 2017, pp. 1-5.

[8] Wei Yuan, Jianhui Wang, Feng Qiu, Chen Chen, Chongqing Kang and Bo Zeng, "Robust Optimization-Based Resilient Distribution Network Planning Against Natural Disasters", in *IEEE Transactions on Smart Grid*, vol. 7, no. 6, pp. 2817-2826, November 2016.

[9] Xu Wang, Mohammad Shahidehpour, Chuanwen Jiang, Zhiyi Li, "Resilience Enhancement Strategies for Power Distribution Network Coupled with Urban Transportation System", in *IEEE Transactions on Smart Grid*, Early Access.

[10] Sfahani, M., Guan, H. and Loo, Y.-C., "Seismic Reliability and Risk Assessment of Structures Based on Fragility Analysis – A Review." *Advances in Structural Engineering*, 18(10), 1653-1669, 2015.

[11] Cimellaro, G. P., Reinhorn, A. M. and Bruneau, M. (2010). "Framework for Analytical Quantification of Disaster Resilience." *Engineering Structures*, 32(11), 3639-3649.

[12] Li, Y., Ahuja, A. and Padgett, J. E. "Review of Methods to Assess, Design for, and Mitigate Multiple Hazards." *Journal of Performance of Constructed Facilities*, 26(1), 104-117, 2011.

[13] Salman, A. M. and Li, Y. "Multihazard Risk Assessment of Electric Power Systems." *Journal of Structural Engineering*, 143(3), 04016198, 2016.

[14] Salman, A. M. and Li, Y. "A Probabilistic Framework for Seismic Risk Assessment of Electric Power Systems." *Procedia Engineering*, 199, 1187-1192, 2017.

[15] Benjamin, J. R. and Cornell, C. A. Probability, Statistics, and Decisions for Civil Engineers, McGraw-Hill, 1970.

[16] Cornell, C. A., Jalayer, F., Hamburger, R. O. and Foutch, D. A. "Probabilistic Basis for 2000 SAC Federal Emergency Management Agency Steel Moment Frame Guidelines." *Journal of Structural Engineering*, 128(4), 526-533, 2002.

[17] D. Bertsekas, Dynamic Programming and Optimal Control, MA, Boston:Athena Scientific, vol. II, pp. 246-253, 2007.

[18] X. Ding, S. L. Smith, C. Belta and D. Rus, "Optimal Control of Markov Decision Processes With Linear Temporal Logic Constraints," in *IEEE Tran. on Automatic Control*, vol. 59, no. 5, pp. 1244-1257, May 2014.